\documentclass[a4, 10pt]{article}
\expandafter\let\csname equation*\endcsname\relax
\expandafter\let\csname endequation*\endcsname\relax
\usepackage{amsmath}

\usepackage{geometry}
\usepackage[justification=justified, labelfont=bf, font={small,singlespacing}]{caption}
\usepackage{color}
\usepackage{graphicx}
\usepackage{epstopdf}
\usepackage{array}
\usepackage{braket}
\usepackage{tikz}
\usepackage{bm}
\usepackage{hyperref}
\hypersetup{colorlinks=true, citecolor=blue, urlcolor=blue}
\usepackage{blindtext}

\newcommand{\sriro}{Sr$_3$Ir$_2$O$_7$}

\newcommand{\address}[1]{\begin{flushleft}{\normalsize{#1}}\end{flushleft}}

\begin{document}
\title{\vspace{-2.0cm}\bf{Critical fluctuations in the spin-orbit Mott insulator Sr$_3$Ir$_2$O$_7$}}
\date{}
\normalsize{
\author{J. G. Vale$^{1,2}$\footnote{Email: {j.vale@ucl.ac.uk}}, S. Boseggia$^{1,3}$, H. C. Walker$^{4,5}$, R. S. Springell$^6$, E. C. Hunter$^7$\footnote{Current address: 
Inorganic Chemistry Laboratory, South Parks Road, Oxford, OX1 3QR, United Kingdom}, R. S. Perry$^{1,8}$, \\ S. P. Collins$^3$, D. F. McMorrow$^1$}
}
\maketitle
\vspace{-0.7cm}

\address{$^1$ London Centre for Nanotechnology and Department of Physics and Astronomy, University College London (UCL), Gower Street, London, WC1E 6BT, United Kingdom}
\vspace{-0.7cm}
\address{$^2$ Laboratory for Quantum Magnetism, \'{E}cole Polytechnique F\'{e}d\'{e}rale de Lausanne (EPFL), CH-1015, Switzerland}
\vspace{-0.7cm}
\address{$^3$ Diamond Light Source, Harwell Science and Innovation Campus, Didcot, Oxfordshire, OX11 0DE, United Kingdom}
\vspace{-0.7cm}
\address{$^4$ Deutsches Elektronen Synchrotron DESY, 22607 Hamburg, Germany}
\vspace{-0.7cm}
\address{$^5$ ISIS Neutron and Muon Source, Science and Technology Facilities Council, Rutherford Appleton Laboratory, Didcot, Oxfordshire, OX11 0QX, United Kingdom}
\vspace{-0.7cm}
\address{$^6$ Interface Analysis Centre, School of Physics, HH Wills Physics Laboratory, University of Bristol, Tyndall Avenue, Bristol BS8 1TL, United Kingdom}
\vspace{-0.7cm}
\address{$^7$ School of Physics and Astronomy, The University of Edinburgh,
James Clerk Maxwell Building, Mayfield Road, Edinburgh EH9 2TT, United Kingdom}
\vspace{-0.7cm}
\address{$^8$ UCL Centre for Materials Discovery, University College London (UCL), Gower Street, London, WC1E 6BT, United Kingdom}

\begin{abstract}
X-ray magnetic critical scattering measurements and specific heat measurements were performed on the perovskite iridate \sriro. We find that the magnetic interactions close to the N\'{e}el temperature $T_N=283.4(2)~\mathrm{K}$ are three-dimensional. This contrasts with previous studies which suggest two-dimensional behaviour like Sr$_2$IrO$_4$.
Violation of the Harris criterion ($d\nu>2$) means that weak disorder becomes relevant. This leads a rounding of the antiferromagnetic phase transition at $T_N$, and modifies the critical exponents relative to the clean system. Specifically, we determine that the critical behaviour of \sriro\ is representative of the diluted 3D Ising universality class.
\end{abstract}

\noindent{\it Keywords\/}: Magnetic critical scattering, iridates, phase transitions 


\section*{Introduction}
The Ruddlesden-Popper series Sr$_{n+1}$Ir$_n$O$_{3n+1}$ of perovskite iridates remain a fruitful arena for study.
Of central importance in these materials is the $j_{\text{eff}}=1/2$ ground state. This state is formed from the interplay of a strong cubic crystal field and spin-orbit interaction on the 5d$^5$ electrons of the Ir$^{4+}$ ions. Weak electron correlations are then sufficient to split the $j_{\text{eff}}=1/2$ band, open an insulating gap, and form a Mott-like state. 

The single-layer compound Sr$_2$IrO$_4$  ($n=1$) exhibits an insulating gap $\Delta E_g \sim 0.4~\text{eV}$, as determined from optical conductivity, resistivity, and scanning tunnelling microscopy (STM) measurements \cite{moon2008, moon2009, okada2013}.
Similarities have been drawn between Sr$_2$IrO$_4$ and La$_2$CuO$_4$ in terms of its structural and magnetic properties, initially leading to claims that Sr$_2$IrO$_4$ is an example of a 2D Heisenberg antiferromagnet on a square lattice (2DHAFSQL) \cite{kim2012_sr214, fujiyama2012, dhital2013}. 
However X-ray magnetic critical scattering \cite{vale2015}, re-examination of previously published resonant inelastic X-ray scattering (RIXS) data, and electron spin resonance (ESR) measurements \cite{bahr2014, bogdanov2015} reveal the presence of a weak easy-plane anisotropy. This anisotropy is caused by pseudodipolar interactions arising fundamentally from the intrinsic spin-orbit interaction, and coupling to the crystal lattice \cite{liu2018, porras2018}.

Meanwhile the bilayer compound \sriro\ marginally retains the $j_{\text{eff}}=1/2$ ground state, and is proximate to a insulator-metal transition (IMT) which can be driven either by injection of carriers \cite{li2013, delatorre2014, hogan2015} or control of the bandwidth \cite{li2013, carter2013, donnerer2016, ding2016}.
The reduced insulating gap compared to the single-layer material -- $\Delta E_g \sim 0.1~\text{eV}$ -- directly manifests from an increase in bandwidth due to the increased dimensionality \cite{moon2008, okada2013}.
Interactions within the bilayer also lead to significant correlations along the $c$-axis, which are not present in Sr$_2$IrO$_4$.
These correlations, combined with significant anisotropic exchange interactions, give rise to G-type antiferromagnetic order below the N\'{e}el temperature $T_N \approx \text{280~K}$, with Ir magnetic moments aligned along the $c$-axis \cite{boseggia_prb2012, boseggia_jpcm2012, dhital2012, fujiyama2012_sr327, kimjw2012}. The anisotropy also results in a large spin gap ($\Delta E_s=85~\text{meV}$) which has been observed by RIXS \cite{kim2012_sr327, moretti2015} and Raman scattering \cite{gretarsson2016}.

Critical scattering studies provide information complementary to that obtainable from the ordered state. For thermally driven transitions in classical systems, issues such as dimensionality, relevant anisotropies, etc., can be addressed by determining the critical exponents both below and above the transition temperature \cite{collins1989}.
Typically one attempts to determine the dimensionality of a material by fitting the magnetisation below the critical temperature $T_{\text{c}}$ to a power law $M\sim (-t)^{\beta}$, where $t=\left(T-T_{\text{c}}\right)/T_{\text{c}}$. The obtained value of the critical exponent $\beta$ can then be compared to the theoretical value for a particular universality class. For a magnetic system these universality classes are governed by the spatial dimensions and spin degrees of freedom.
A number of thermodynamic parameters, including the susceptibility, correlation length, and magnetic specific heat, exhibit similar behaviour -- with their own critical exponents -- in the vicinity of the critical point. 
Previous estimates for the critical exponent $\beta$ \cite{dhital2013, dhital2012, miyazaki2015} for \sriro\ implied that the magnetic interactions were two-dimensional at $T_N$. 
However, these estimates for the critical exponents were obtained using local probes over a wide range of temperatures, and are insensitive to subtle effects driven by weak anisotropies close to the transition temperature.

We performed critical scattering and specific heat measurements on \sriro\, in order to precisely determine the spin and lattice dimensionality in the vicinity of the N\'{e}el temperature. In contrast with the previous results, we find that the critical fluctuations are three-dimensional in nature. Fundamentally this is a consequence of the significant interlayer coupling and intrinsic anisotropy, giving rise to significant differences in the observed behaviour compared to Sr$_2$IrO$_4$.

\section*{Methods}

\begin{figure}[t]
\centering
\includegraphics[scale=1]{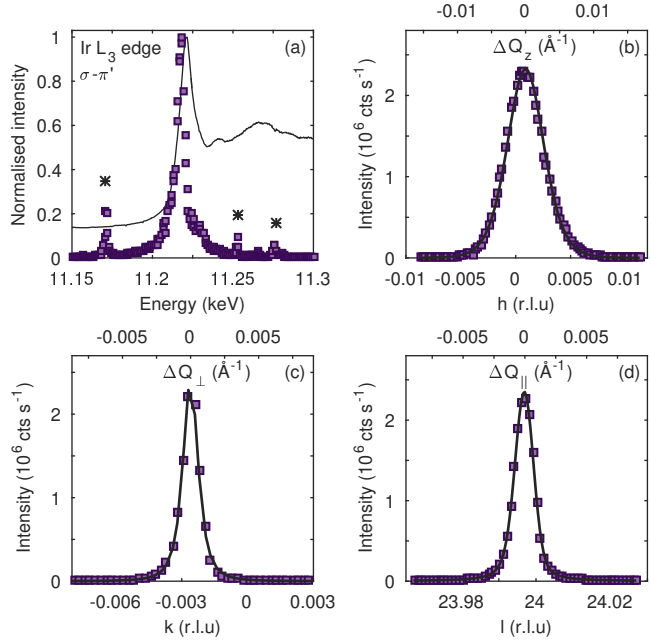}
\caption{(a): Energy scan of the $\left(\frac{1}{2},\frac{1}{2},24\right)$ magnetic Bragg peak with $\sigma$--$\pi'$ polarisation (symbols). Overlaid is the total fluorescence yield (TFY) from the sample (solid line). Peaks marked with asterisks result from multiple scattering. (b)--(d): Reciprocal space scans of the $\left(0,\,0,\,24\right)$ peak at 200~K, which was used to represent the resolution function. Nearby structural peaks exhibited similar behaviour. Added are the best fit (solid lines) of the data to a Voigt function.}
\label{Sr327_fluo_res}
\end{figure}

The critical scattering experiments were performed on beamline I16, Diamond Light Source. A single crystal of \sriro\ (dimensions 0.5$\times$0.5$\times$0.3mm$^3$) was flux grown from the phase-pure polycrystalline compound using techniques described elsewhere \cite{li2013}, and attached to the copper sample mount of a He closed-cycle refrigerator (Displex 4K).
This was in turn mounted on a six-circle diffractometer configured to operate in a vertical scattering geometry. The energy of the incident photon beam was set to 11.218~keV, just below the $L_3$ edge of iridium, a value found to maximise the intensity of the X-ray resonant magnetic scattering (Fig.~\ref{Sr327_fluo_res}a). The incident beam size was determined to be
$200\times 20~\mu\mathrm{m}^2$ (H$\times$V). The polarization of the scattered X-rays was determined by using a pyrolytic graphite $(0,0,8)$ crystal analyser mounted on the detector arm. The temperature was measured to a precision of $\pm$0.01~K via a thermocouple secured to the sample mount by Teflon tape. 
The wavevector resolution of the instrument, including the effects of sample mosaic, was determined by mapping Bragg peaks in reciprocal space.\footnote{The profile of the resolution function is dictated by a number of factors. These include the shape of the source, choice of monochromator, use of collimating slits and focussing mirrors \cite{cowley1987}.} This was found to be typically (at FWHM) 1.3$\times 10^{-3}$ \AA$^{-1}$ and 1.5$\times 10^{-3}$ \AA$^{-1}$ perpendicular and parallel to $\mathbf{Q}$ in the scattering plane respectively, and 6.3$\times 10^{-3}$ \AA$^{-1}$ out of the plane (Fig.~\ref{Sr327_fluo_res}b--d).

\begin{figure}
\centering
\includegraphics[scale=1]{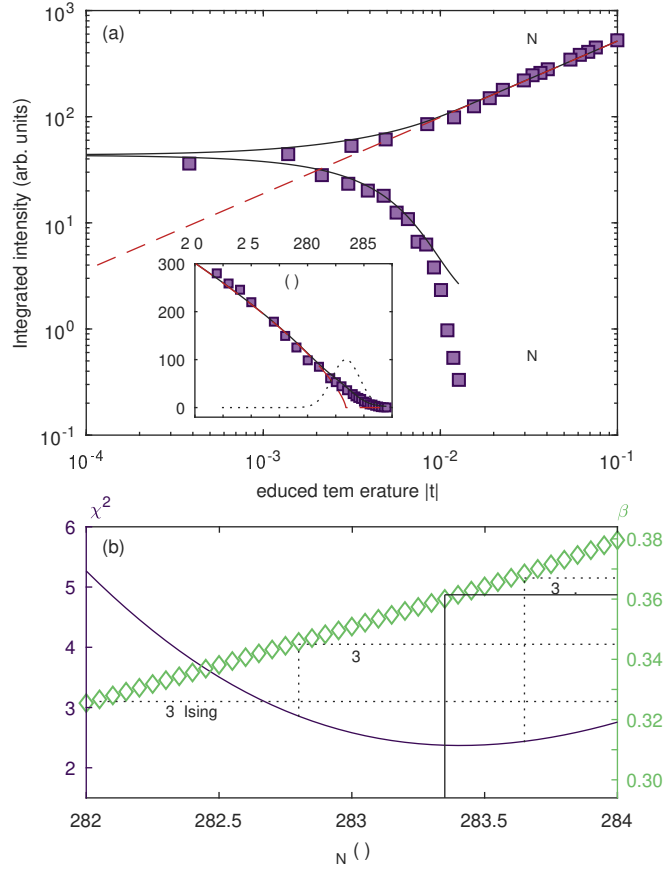}
\caption{(a): Integrated intensity of $\left(\frac{1}{2},\frac{1}{2},24\right)$ magnetic Bragg peak in terms of the absolute reduced temperature $\vert t \vert = \vert 1 - T/T_N \vert$, as obtained from $\theta$--$2\theta$ scans. Solid line: Best fit to power law convoluted with a Gaussian distribution of $T_N$ with FWHM $2.7(1)~\mathrm{K}$. Dashed line: Corresponding power law assuming single value of $T_N$.
Inset is the same data plotted on linear axes in the vicinity of $T_N$, highlighting the rounding of the transition. Dotted line: Gaussian distribution of $T_N$.
(b): Variation of $\chi^2$ (solid line) and $\beta$ (diamonds) as a function of $T_N$. The minimum of the $\chi^2$ surface occurs for
$T_N=283.4(2)~\mathrm{K}$ and $\beta=0.361(8)$, which is consistent with the theoretical value for 3D Heisenberg interactions.
}
\label{Sr327_op}
\end{figure}

\section*{Results}
The first objective was to determine the N\'{e}el temperature and the critical exponent $\beta$.
Scans of the $\left(\frac{1}{2},\,\frac{1}{2},\,24\right)$ magnetic Bragg peak were performed parallel to the scattering wavevector $Q$ ($\theta$--$2\theta$ scans) as a function of temperature. At each temperature the observed peak was fitted to a Lorentzian squared lineshape convoluted with the experimental resolution function, which was found to best represent the data.
The resulting integrated intensity is plotted in Fig.~\ref{Sr327_op}(a). One can see that the intensity decreases continuously as a function of increasing temperature -- as would be expected for a second-order magnetic phase transition -- going towards zero around 290~K.
Unlike a perfect second-order phase transition, however, the transition is not especially sharp, but exhibits a degree of rounding. Rounding of the transition can occur due to sample inhomogeneity or defects within the scattering volume, since different areas of the sample will have slightly different ordering temperatures (random $T_c$ disorder).

Consequently the integrated intensity was fitted between 250 and 290~K (corresponding to a reduced temperature $-\text{0.12} \leq t  \leq  \text{0.023}$)  with a power law: $I_M \propto{} (-t)^{2\beta}$. This was then convoluted with a Gaussian distribution of transition temperatures with FWHM $\Gamma$ to model the effect of disorder. Note that such a functional form is no longer linear when plotted on double logarithmic axes, unlike a conventional power law. 
It was found that the best fit to the data was obtained with $T_N=283.4(2)~\mathrm{K}$, $\beta = 0.361(8)$ and $\Gamma=2.7(1)~\mathrm{K}$. The deviation of the model from the experimental data at high temperatures is due to critical scattering in the paramagnetic phase. This is characterised by a different exponent, and shall be discussed in more detail later in the manuscript.
Our value for $T_N$ is in good agreement with that obtained from bulk magnetisation measurements and neutron powder diffraction \cite{dhital2012}.
On the other hand, the value of $\beta$ differs significantly from previous neutron scattering \{$\beta=0.25$ \cite{dhital2012}, $0.20(2)$ \cite{dhital2013}\} and $\mu$SR \{$\beta=0.143(3)$ \cite{miyazaki2015}\} measurements on the same material. We suggest the discrepancies arise because in the previous works, $\beta$ was determined from power law fits which included a significant number of data points far from $T_N$. Strictly speaking, power law scaling for thermodynamic parameters is only exact precisely at $T_N$. The results presented here provide a more reliable estimate for $T_N$ and $\beta$.
We note that $\beta=0.361(8)$ is consistent with the theoretical value for a 3D Heisenberg model ($\beta=0.367$). 


\begin{figure}[t]
\centering
\includegraphics[scale=1]{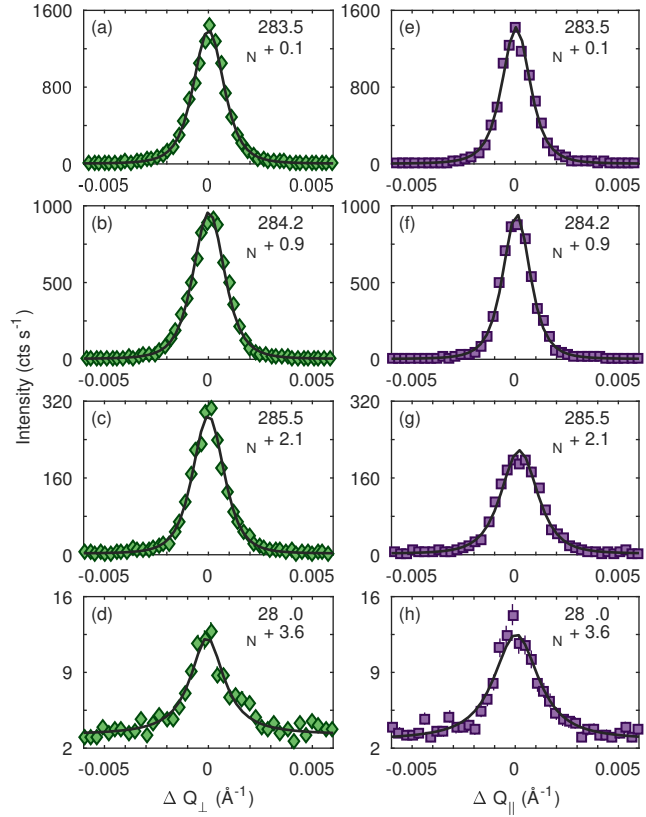}
\caption{Evolution of the $\left(\frac{1}{2},\frac{1}{2},24\right)$ magnetic Bragg peak above $T_N$ in the $k$- and $l$-directions (left and right columns respectively). The critical scattering is nearly isotropic. Solid lines are best fit to a Lorentzian-squared function convoluted with the instrumental resolution function.}
\label{Sr327_critcomp}
\end{figure}

The variation in lattice dimensionality between Sr$_2$IrO$_4$ and \sriro\ should also manifest in differences in the critical phenomena \emph{above} the N\'{e}el temperature. 
The same magnetic Bragg peak was followed out to 288~K ($T_N + 5.2~\mathrm{K}$) for both the in-plane ($\mathbf{Q} \parallel k$) and out-of-plane ($\mathbf{Q} \parallel l$) directions, with a comparison plotted in Fig.~\ref{Sr327_critcomp}. Compared to Sr$_2$IrO$_4$ \cite{vale2015}, the critical scattering in \sriro\ appears practically isotropic, and decays much more quickly with temperature.
This would be expected for systems where 3D interactions are important. Remember that the N\'{e}el temperature is associated with the onset of long-ranged three-dimensional antiferromagnetic order. The intensity previously observed for Sr$_2$IrO$_4$ above $T_N$ only arises as a result of dominant 2D correlations in-plane which have a much larger energy scale than those out-of-plane. If the dominant magnetic interactions are three-dimensional, then one would expect both the in-plane and out-of-plane interactions to behave in a similar manner (to first order). In this sense \sriro\ is more `conventional'. Consequently the correlation length $\xi$ and equal-time structure factor $S_0$ should be expected to follow the simple power laws: $\xi \sim t^{-\nu}$ and $S_0 \sim t^{-\gamma}$.





Within the Ornstein-Zernike approximation, the spin-spin correlation function decays exponentially as a function of distance. In reciprocal space, this leads to a Lorentzian functional form in the vicinity of the critical temperature. 
X-ray magnetic critical scattering has the advantage of satisfying the condition $\hbar\omega<E_i$, where $\hbar\omega$ refers to the energy of the critical fluctuations. Moreover, one typically collects data by collecting all scattered X-ray photons without regard to their energy.
Thus the static approximation holds, and the resulting scattering cross-section for the critical fluctuations should also have a Lorentzian functional form. The amplitude of this Lorentzian $S_0$ is proportional to the staggered susceptibility $\chi_0$, whilst the inverse half width at half maximum (HWHM) $\Gamma^{-1}$ corresponds to the correlation length $\xi$.

\begin{figure}[t]
\centering
\includegraphics[scale=1]{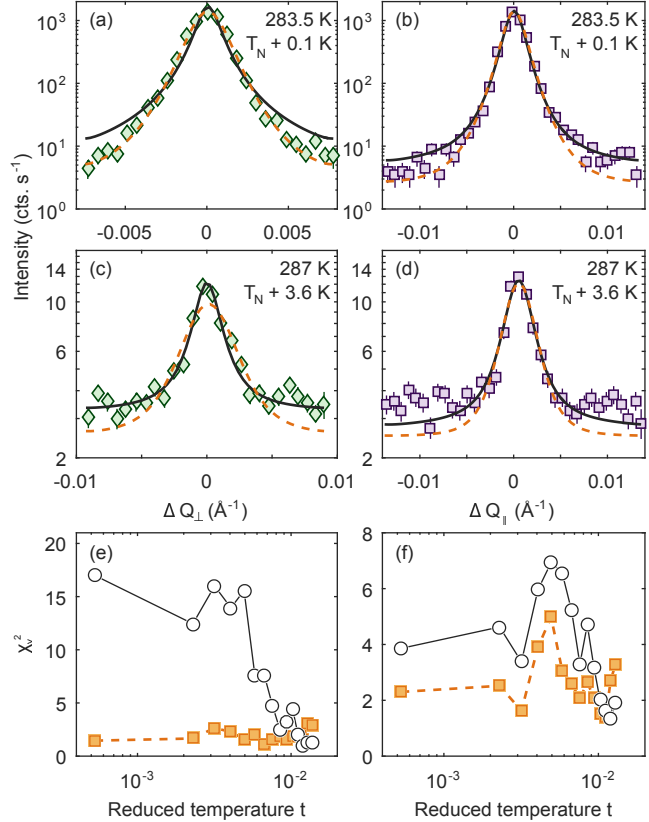}
\caption{Lineshape comparison of fits to critical scattering for the in-plane (a,c) and out-of-plane (b,d) directions. Solid black (dashed orange) line indicates a fit using a Lorentzian (squared) scattering function. Results are displayed on a logarithmic scale to highlight differences in the peak tails. The displayed data has been rebinned in the momentum direction $\Delta Q_{\perp}$ for clarity;
all fits were performed on the full dataset. Bottom panels: Variation of reduced $\chi^2$ as a function of reduced temperature $t=T/T_N - 1$, for scans along $Q_{\perp}$ (e) and $Q_{\parallel}$ (f). Open circles (filled squares) indicate fits to a Lorentzian (squared) scattering function.}
\label{Sr327_lineshape_comp}
\end{figure}

Yet upon fitting the lineshapes of the magnetic Bragg peaks above $T_N$, one finds that a Lorentzian squared function (convoluted with the instrumental resolution function) is consistently more representative of the data than a Lorentzian function (Fig.~\ref{Sr327_lineshape_comp}). Note that the background for each fit was fixed based on scattering well away from the magnetic Bragg peak.
However the discrepancies in the lineshape are almost entirely in the tails of the peak; the fitted values of the peak width are almost identical for both lineshapes. Furthermore for data collected along $Q_{\parallel}$, the differences between the Lorentzian and Lorentzian squared functional forms are small. Given that a Lorentzian squared form is a better fit to both the $Q_{\perp}$ and $Q_{\parallel}$ data, this is what has been used in the subsequent analysis. 
\footnote{We remark that the HWHM of any Lorentzian raised to a power $n$ is obtained by dividing the corresponding width parameter $\Gamma$ by a factor of $\left[2^{1/n}-1\right]^{1/2}$. All data displayed has been corrected in this way.}


\begin{figure}[h]
\centering
\includegraphics[scale=1]{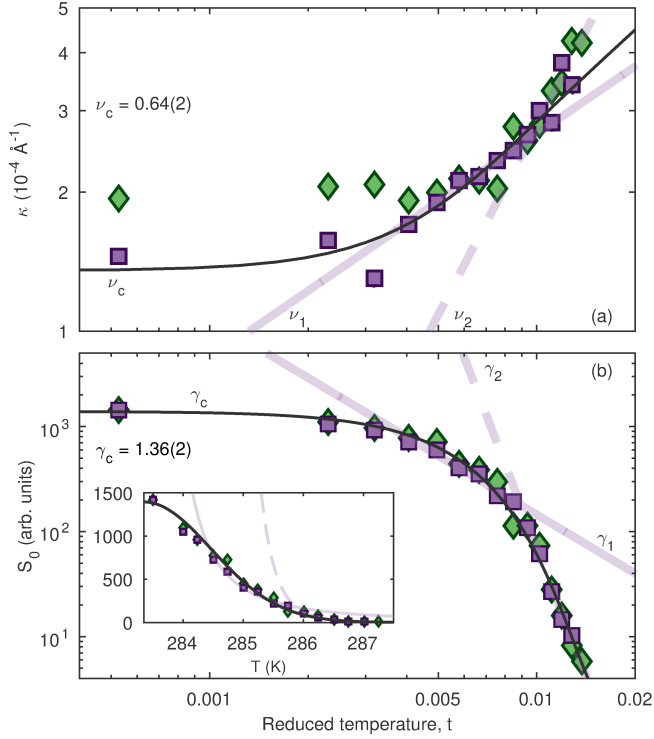}
\caption{Inverse correlation length (a) and intensity (b) of $\left(\frac{1}{2},\frac{1}{2},24\right)$ magnetic Bragg peak as a function of reduced temperature $t\!=\!T/T_N-1$. 
Green diamonds: in-plane direction. Purple squares: out-of-plane direction.
Solid and dashed purple lines are fits of the out-of-plane data to the relevant power law for two different temperature regions as described in the main text: $\nu_1=0.49(4), \gamma_1=1.86(3)$; $\nu_2=1.4(1), \gamma_2=8.2(3)$. Solid black lines are the best fit to a power law convolved with a Gaussian distribution of $T_N$ [FWHM $2.7(1)~\mathrm{K}$]: $\nu_c = 0.64(2)$, $\gamma_c=1.36(2)$. Inset in (b) is the same as the main panel, only plotted on linear axes. 
}
\label{Sr327_aboveTN}
\end{figure}
The fitted values of the inverse correlation length $\kappa$ and the peak amplitude $S_0$ are plotted in Figure \ref{Sr327_aboveTN} for data collected along $Q{_{\parallel}}$ and $Q_{\perp}$. The two datasets appear isotropic, with the exception of the correlation length for $t<\text{0.005}$. Some of this discrepancy may be due to the deviation of the observed lineshape from an ideal Lorentzian. 
The precise form of the spin-spin correlation function is dependent on the critical exponent $\eta$, which is defined at $T=T_{\text{N}}$ and becomes increasingly relevant close to that limit. However for three-dimensional systems, $\eta \approx \text{0}$ and so this is likely to be a minor effect.
An alternative is that two components may contribute to the observed scattering close to $T_{\text{N}}$. The second component could, for example, result from defect mediated scattering, which experimentally \cite{andrews1986, mcmorrow1990, thurston1994, cowley1996, huennefeld2002} and theoretically \cite{altarelli1995, papoular1997} has been shown to have a Lorentzian squared lineshape.
However it was not possible to unambiguously resolve more than one component in the data. As the $Q_{\parallel}$ data shows qualitatively more ideal behaviour, quantitative analysis shall be restricted to this dataset from now on.

The saturation of $\kappa$ and $S_0$ for $t<\text{0.005}$ suggests that a single power law is insufficient to fully describe the data. Two distinct temperature regions can be observed in which the data appears linear on logarithmic axes ($\text{0.003} < t < \text{0.01}$ and $\text{0.007} < t < \text{0.015}$ respectively). However upon performing simple power law fits ($\kappa \sim t^{\nu}$, $S_0 \sim t^{-\gamma}$) to the data in these regions, one finds that the critical exponents associated with these fits do not correspond to the theoretical values associated with any conventional universality class. 
Recall that the order parameter data showed a rounding of the phase transition, which was best described by the convolution of a power law $I \sim -t^{2\beta}$ with a Gaussian distribution of $T_{\text{N}}$. Yet the simple power law fits to $\kappa$ and $S_0$ assume a single value of $T_{\text{N}}$. 
Convolution of a single power law with a Gaussian distribution of $T_{\text{N}}$ (of the same width as that used for the order parameter data) provides a much better description of the data. 


The critical exponents arising from the order parameter, inverse correlation length, and susceptibility are all consistent with three-dimensional magnetic interactions (Table \ref{exponent_table}). However there is some discrepancy in the spin dimensionality that these exponents represent. 
This may (in part) be due to a slight underestimate of $\nu$, as a consequence of the complications with the lineshape close to $T_N$. We note that the critical exponent $\eta$, obtained using the scaling relation $\gamma=\nu(2-\eta)$, is considerably smaller than the theoretical value for any 3D spin model. The same goes for the experimental ratio $\beta/\nu$, which only marginally agrees with theory. Assuming that ideal scaling is satisfied, and that $\beta$ and $\gamma$ are correct, then this would imply that in fact $\nu=0.69(1)$. 

\begin{figure}
\centering
\includegraphics[scale=1]{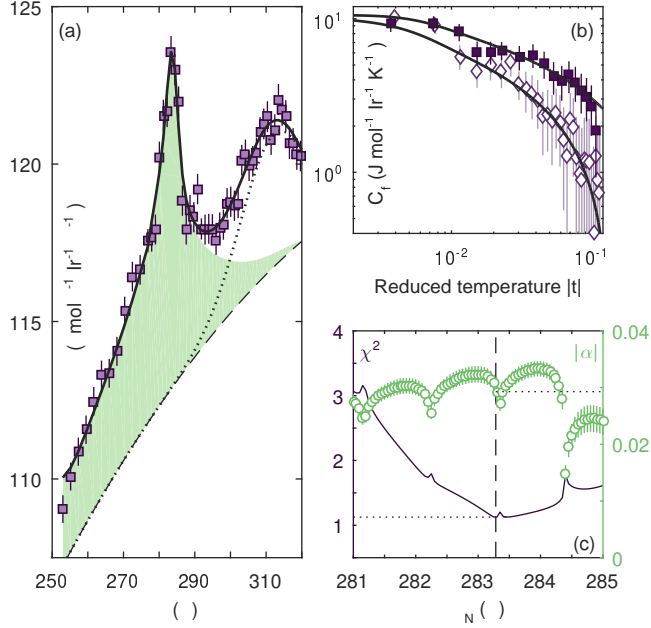}
\caption{(a): Specific heat data collected on \sriro\ with zero applied magnetic field. Solid line is the best fit to the data as described in the main text (Eqn.~\ref{Cp_eqn}). Dashed and dotted lines represent a polynomial phonon background $C_{latt}$ and 310~K peak $C_{peak}$ respectively. (b): Fluctuation contribution to heat capacity $C_f$ with respect to reduced temperature $|t|$. Filled (open) symbols refer to data collected below (above) $T_N^C$ respectively. Note that the contribution from the anomalous peak at 310~K was subtracted from the experimental data. (c): Variation of $\chi^2$ (solid line) and $|\alpha|$ (diamonds) as a function of fixed $T_N$. The best fit to the data was obtained using $T_N^C=\text{283.3(5)~K}$, $|\alpha|=0.03(1)$ (dotted line).}
\label{Sr327_Cp}
\end{figure}

\begin{table*}[ht!]
\centering
\scriptsize
\begin{tabular}{p{2.0cm}|lllll|lll|l}
 & $\beta$ & $\nu$  & $\gamma$ & $\alpha$ & $\eta$  & $\alpha+2\beta+\gamma$ & $\alpha+d\nu$ & $\beta/\nu$ & $U_0$\\
\hline
3D Ising & 0.327(1) & 0.630(1) & 1.237(1) & 0.110(1) & 0.036(1)  & 2.000(2) & 2.000(2) & 0.518(1) & 0.53(3)\\
3D XY & 0.345 & 0.669(7) & 1.316(9) & -0.01 & 0.03  & 1.996(9) & 2.00(3) & 0.516(6) & 1.06(3) \\
3D Heisenberg & 0.367 & 0.707(3) & 1.388(3) & -0.121 & 0.037  & 2.001(3) & 2.00(1) & 0.519(2) & 1.52(3)\\
\hline
3D dilute Ising & 0.354(2) & 0.683(3) & 1.342(6) & -0.049(9) & 0.035(2) & 2.00(1) & 2.00(1) & 0.518(4) & 1.6(3)\\
\hline
\hline
\sriro\ (raw) & 0.361(8) & 0.64(2) & 1.36(2) & -0.03(1) & -0.13(1)  & 2.05(3) & 1.89(6) & 0.56(4) & 1.0(1) \\
\sriro\  &  & 0.69(1) &  &  & 0.03(1)  &  & 2.04(4) & 0.52(2) & \\
(adjusted $\nu$) & & & & & & & & \\
\end{tabular}
\caption{Comparison of critical exponents and universal amplitude ratio $U_0$ obtained experimentally for \sriro\ with theoretical results summarised by various authors \cite{collins1989, ballesteros1998, pelissetto2002, calabrese2003_1, calabrese2003_2, hasenbusch2007}. The experimental value for $\eta$ has been calculated using the Widom scaling relation $\gamma=\nu(2-\eta)$. For reference the expected scaling laws have also been added.}
\label{exponent_table}
\end{table*}
Further thermodynamic parameters also exhibit power-law behaviour in the vicinity of the critical temperature. Of these, the most easily accessible experimentally is the specific heat.
Specific heat data published by Nagai \cite{nagai2007} shows a cusp around 280~K, which corresponds to a second-order phase transition. The authors proposed a magnetic origin for the feature, given that it coincides with $T_N$ determined from susceptibility data.
We obtained specific heat data on a single crystal of \sriro\ from the same batch as the sample used for the critical scattering measurements. Our data (plotted as inset of Fig.~\ref{Sr327_Cp}) is consistent with the previously published results. Note that an additional broad peak can be observed at 310~K. The origin of this peak is unclear at present, but may be related to some degree of two-dimensional magnetic fluctuations \cite{sengupta2003}. We comment that a similar feature (albeit much weaker) can be seen upon careful examination of the data in Ref.~\cite{nagai2007}. Our focus remains upon the sharp peak at 283~K.

In general the specific heat $C_{tot}$ is the sum of two contributions: a non-singular component $C_{latt}$ (expected to be dominated by the lattice at the temperatures studied), and the fluctuation specific heat $C_f$. We fitted the experimental data (including the unexplained peak at $T_p=\text{310~K}$) using the expression:
\begin{align}\label{Cp_eqn}
C_{tot} = \underbrace{aT+bT^2}_\mathrm{{C_{latt}}} + \underbrace{C_0 - A_{\pm}(\pm t)^{-\alpha}}_\mathrm{C_f} + \underbrace{A_{p}e^{-\frac{(T-T_{p})^2}{2\sigma_p^2}}}_\mathrm{C_{peak}},
\end{align}
with the positive sign required if the reduced temperature $t=T/T_N - 1 > 0$, and vice versa.
This function was then convolved with a Gaussian distribution, in order to account for the rounding of the phase transition. The distribution of $T_N$ was assumed to have an identical FWHM ($\Gamma=\text{2.7(1)~K}$) as that used for the critical scattering data. Treating $\Gamma$ as a free parameter, or using a simple Debye model (with $\Theta_D = 380~\text{K}$) to describe $C_{latt}$, did not alter our results appreciably.

The best fit to the data (Fig.~\ref{Sr327_Cp}a) was obtained with $T_N^C=\text{283.3(1)~K}$, $U_0=A_{+}/A_{-}=1.03(1)$, and $\alpha=-0.028(2)$. However, we note that the $\chi^2$ surface (Fig.~\ref{Sr327_Cp}c) exhibits a number of sharp anomalies, which coincide with the locations of experimental data points. This is due to limited data for $|t|<1\times10^{-2}$. Nevertheless, the general trend can be clearly observed. We thus propose the following, more conservative, estimates for the relevant parameters: $T_N^C=\text{283.3(5)~K}$, $\alpha=-0.03(1)$, $U_0=1.0(1)$.

The value of the N\'{e}el temperature obtained from the specific heat measurements is in good agreement with that obtained from critical scattering. Moreover, the determined values of $\alpha$ and $U_0$ are intermediate between the 3D XY and 3D Heisenberg universality classes, again consistent with the results presented earlier in this manuscript.


\section*{Discussion}
There is, however, one point that has only been briefly touched upon thus far: the effect of disorder. 
The Harris criterion ($d\nu>2$) reflects the stability of a clean critical point against the effect of disorder. If the inequality is satisfied, then the critical point is stable against disorder. Weak disorder decreases under coarse graining, and becomes unimportant on large length scales.
On the other hand, if $d\nu<2$, the converse is true, and disorder becomes a relevant perturbation. A new universality class results, with exponents which now satisfy $d\nu>2$. The global phase transition can also become smeared, as while the global magnetization develops gradually, rare regions order independently.
For instance, a clean 3D Heisenberg system [$\nu = 0.707(3)$] is stable against weak disorder, while a 3D Ising system [$\nu = 0.6301(4)$] is not. Consequently a dirty 3D Ising system is characterised by critical exponents which differ from the clean case.
One theoretical description of a dirty 3D Ising system is given by the following Hamiltonian:
\begin{equation}
\mathcal{H} = \sum_{i,j,\alpha,\beta} J_{ij}^{\alpha\beta}\bm{\epsilon}_i \bm{\epsilon}_j S_i^{\alpha}S_j^{\beta}.
\end{equation}
Here the $\bm{\epsilon}$'s are quenched, uncorrelated random variables, chosen to be 1 with probability $p$ (the spin concentration), or 0 with probability $1-p$ (the impurity concentration, or spin dilution). In the literature it is also known as the three-dimensional \emph{diluted} Ising model (3DDI). It has been found that the critical exponents are independent of disorder above the percolation limit ($p_c=0.31$), and are clearly distinct from the clean 3D Ising case. Note that the 3D Ising model with random-$T_c$ disorder has also been shown to lie in the same universality class \cite{vojta2006}

The theoretical results for the 3DDI model were compared with the critical behaviour observed for \sriro. Our experimental values for the critical exponents (after adjusting for $\nu$ as described above) are in good agreement with the calculated ones \cite{ballesteros1998, hasenbusch2007}. 
It is possible to go further and compare the theoretical transition temperature with the experimental value of $T_N$.
Ballesteros \emph{et al.}~performed a Monte Carlo finite size scaling (FSS) analysis of the 3DDI model, and calculated the critical temperature $T_c$ for various levels of $p$. Extrapolating their data to the limit $p\!\rightarrow\!1$ -- since we expect the level of disorder in \sriro\ to be small -- we find that $T_c (p\rightarrow 1) \sim 0.21 \tilde{J}$, where $\tilde{J}$ is some effective next-nearest neighbour coupling strength.
The Hamiltonian for \sriro\ includes a number of further neighbour and anisotropic exchange terms, the magnitude and uncertainty of which have been previously determined by resonant inelastic X-ray scattering \cite{kim2012_sr327}. Through evaluating the sum: $\tilde{J} = \sum_i z_i J_i$, where $J_i$ are the individual coupling parameters (including anisotropies), and $z_i$ the number of neighbours, we obtain $\tilde{J}=\text{128(18)~meV}$ for \sriro. This corresponds to a transition temperature of $T_c = \text{320(40)~K}$, which is in agreement with the experimental value of $T_N=\text{283.4(2)~K}$.
Both observations confirm that the critical behaviour in \sriro\ is consistent with a 3D Ising model with quenched disorder.

\section*{Conclusion}
We have studied the critical fluctuations in the spin-orbit Mott insulator \sriro\ via resonant elastic x-ray scattering and bulk specific heat measurements. In contrast with previous studies, we determine that the magnetic fluctuations are three-dimensional in the vicinity of the critical temperature $T_N$. Weak disorder leads both to a smearing of the global phase transition, and deviation of the critical exponents from those expected for a clean 3D Ising system. We further establish that the observed behaviour can be well described by the diluted 3D Ising universality class.
Our observations are consistent with the significant uniaxial anisotropy present in the magnetic ground state. 

\section*{Acknowledgements}
The authors thank Diamond Light Source for allocation of beamtime under proposal MT7798.  Work in London was supported by the EPSRC (Grants No. EP/N027671/1, EP/N034872/1). J.~G.~V. would like to thank UCL and EPFL for financial support through a UCL Impact award.

\bibliographystyle{unsrt}

\end{document}